# ZYELL-NCTU NetTraffic-1.0: A Large-Scale Dataset for Real-World Network Anomaly Detection

Lei Chen, Shao-En Weng, Chu-Jun Peng, Hong-Han Shuai, and Wen-Huang Cheng
National Yang Ming Chiao Tung University, Taiwan
{lami860516.eic08g, anita4213.eed05, greenlight2124.eecs06, hhshuai, whcheng}@nctu.edu.tw

*Abstract*— Network security has been an active research topic for long. One critical issue is improving the anomaly detection capability of intrusion detection systems (IDSs), such as firewalls. However, existing network anomaly datasets are out of date (i.e., being collected many years ago) or IP-anonymized, making the data characteristics differ from today's network. Therefore, this work introduces a new, large-scale, and real-world dataset, ZYELL-NCTU NetTraffic-1.0, which is collected from the raw output of firewalls in a real network, with the objective to advance the development of network security researches.

## I. INTRODUCTION

Network security has been a critical issue for long with the enormous growth of computer network usage and billions of applications running on top of it. Regardless of personal or enterprise users, hackers today have found more opportunities to hack crucial information. Therefore, it becomes more important for intrusion detection systems (IDSs) to reliably detect attacks and anomalies in networks. Developing robust detection techniques on IDSs to defend and protect against ever-growing malicious behaviors is imperative.

On the other hand, network anomaly detection (NAD) is an attempt to solve these problems. NAD identifies anomalous network traffic by observing past given data over time. According to the detection principles, there are two policy types, signature-based and rule-based. The signature-based policy distinguishes the traffic between normal and abnormal by manually adding patterns to the white list. The rule-based method, however, induces a representative and condense rule from hundreds or thousands of signatures with AI. Due to the diversity and rapidly increasing number of malicious behaviors, machine learning methods, particularly deep learning, have received much attention to deal with the anomalous problems, but it is still an open challenge since real-world network data are large-scale, label-lacking, and class-imbalanced (i.e., anomalies only typically occur 0.001-1% of the time).

Most of the existing network datasets are not meeting the real-world conditions or outdated from modern networks, such as 1998 and 1999 DARPA intrusion detection datasets, KDD'99, Kyoto 2006+, and ISCX2012 [1]-[5]. Particularly, some are IP-anonymized due to privacy, for example, CAIDA OC48 [6] and LBNL [7]. However, removing features like source and destination information will not truly reflect the real network flows. Moreover, lots of IDSs nowadays like firewalls that are the most common IDS products, are still based on expert-defined rules. How to effectively improve the performance of these products is important.

TABLE I
STATISTICS OF THE ZYELL-NCTU NETTRAFFIC-1.0 DATASET

| Normal or Attack Types | # of instances (% of total instances) | # of training data instances (% of training data instances) | # of testing data instances (% of testing data instances) |
|---|---|---|---|
| Normal | 22181694 (98.445) | 8920477 (96.527) | 13261217 (99.779) |
| DDOS-smurf | 5877 (0.026) | 2344 (0.0254) | 3533 (0.027) |
| Probing-IP sweep | 266034 (1.181) | 240524 (2.603) | 25510 (0.192) |
| Probing-Port sweep | 77554 (0.344) | 77289 (0.836) | 265 (0.002) |
| Probing-Nmap | 834 (0.004) | 829 (0.009) | 5 (0.0000376) |
| Total | **22181694** | **9241463** | **13290530** |

In this paper, a joint research task force (National Yang Ming Chiao Tung University and Zyell Solutions of Zyxel Group) introduces a new, million-scale, and real-world firewall system log dataset, namely ZYELL-NCTU NetTraffic-1.0. Instead of being built upon virtual or experimental environments, our dataset is under real networks that include daily network traffic. It can be used to develop solutions to network threats including the distributed denial-of-service attack (DDoS) and probe-response attack (Probing) for firewalls. Through this dataset, we hope to inspire solutions across academic and industrial communities to help advance the field of network security.

## II. DATASET DESCRIPTION

### A. Data Collection

The data are time-series traffic records captured by real firewalls and the total number of collected logs is about 22.5 million. TABLE I gives the statistics of records in the training and testing set. Each log record is a network connection session (or a flow). Attacks is issued randomly with the hping3 and nmap tools. The attack interval varies from three minutes to four hours. Considering the characteristics of the time series, we also record twelve statistical features based on IP address or port of source/destination, as shown in TABLE II. The hyper-parameters are expert-selected. Here, we choose N for 100, T for 3, T' for 600. Final traffic logs contain 22 features, including log collection time, source/destination IP addresses and ports, inbound/outbound traffic counts (bytes), connection duration (seconds), protocol and app type, and the extending features.

TABLE II
ADDITIONAL STATISTICAL FEATURES

| Feature Name | Meaning |
|---|---|
| cnt_dst | For the same source IP address, the number of unique destination IP addresses within the network in the last T seconds. |
| cnt_src | For the same destination IP address, the number of unique source IP addresses within the network in the last T seconds. |
| cnt_serv_src | Number of connections from the source IP to a same destination port in the last T seconds. |
| cnt_serv_dst | Number of connections from the destination IP to a same source port in the last T seconds. |
| * slow[a] | * last T' seconds.[a] |
| * conn[a] | * last N connections.[a] |

[a] * represents the four basic types of features (e.g., cnt_dst) are recorded but with a different hyper-parameter value (i.e., T' and N, respectively).

TABLE III
THE PERFORMANCE SUMMARY ON TESTING SET WITH DIFFERENT MACHINE LEARNING ALGORITHMS

| Model | Accuracy | Precision | Recall | F1 score |
|---|---|---|---|---|
| KNN | 0.4718 | 0.3030 | 0.4718 | 0.3188 |
| Random Forest | 0.6136 | 0.3474 | 0.6136 | 0.3635 |
| QDA | 0.5583 | 0.3995 | 0.5583 | 0.3836 |
| ID3 | 0.5375 | 0.2369 | 0.5375 | 0.2496 |

The data is post-processed from the raw outputs of the firewalls. We only select the traffic statistic part in the firewall output logs and semi-automatically label the data as either normal or an exact attack type. The proportion of anomalies is about 1.5%. It is important to note that the probability distribution of testing data is not the same as training data.

### B. Testbed Architecture

As shown in Fig.1, the network infrastructure can be simply divided into an attack system and victim systems. The attack system, including all attackers' computers, will go through WIFI first, then to a switch. Each USG (Unified Security Gateway) device protects networks with industry-leading firewall, Anti-Malware/Virus, IDP and several other functionalities. The architecture is under ZYELL's environment.

### C. Attack Profiles and Scenarios

We simulate four sub-attacks of DDoS and Probing profiles. DDoS is an attack trying to shut down traffic flow to and from the target system. Probing tries to explore the client's activities and to get information from the network. We select smurf attack, one form of DDoS, and three attacks in probing. The definition of simulated attacks is described as follows:

***DDoS-smurf***: Smurf is similar to ping flood, which sends lots of request packets to crash the target. However, it boosts its damage by exploiting characteristics of broadcast networks.

***Probing-Port sweep***: A port sweep looks for wide ports to determine the services which are supported during a period on a single specific host.

***Probing-IP sweep***: The IP sweep attack aims to find live hosts. It performs either a port sweep or ping on multiple host addresses. The attackers send ICMP requests to multiple destination addresses to find out the replied hosts, once the target replied, its IP address is revealed to the attackers.

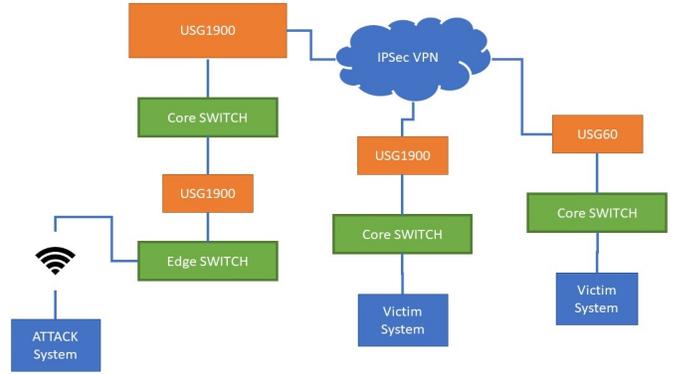

Fig. 1. The testbed architecture to collect syslog of firewalls.

***Probing-Nmap***: The nmap attack maps the network with nmap tool. Here we define nmap attack as ping scan only.

We also conduct some machine learning approaches on the dataset and the results are shown in TABLE III (Here we use macro average and class balanced accuracy). It can be found that no matter the traditional machine learning approaches or basic deep learning models cannot achieve a high performance, and we believe that this dataset is enough challenging to be the new benchmark of network security so as to help narrow the gap between researches and real-world applications.

### III. CONCLUSION

In this paper, we introduce a new million-scale dataset, ZYELL-NCTU NetTraffic-1.0 dataset, which serves to provide attack scenarios from firewall logs. The features of this dataset are seriously imbalanced and truly real-world that built on ZYELL's networks. We believe it can be applied as a new benchmark to help either researchers or industries to improve NAD methods on IDS systems.